\PassOptionsToPackage{unicode}{hyperref}
\PassOptionsToPackage{hyphens}{url}
\PassOptionsToPackage{dvipsnames,svgnames,x11names}{xcolor}
\documentclass[12pt]{article}

\usepackage{amsthm}
\theoremstyle{remark}
\newtheorem{remark}{Remark}

\usepackage{amsmath,amssymb}
\usepackage{iftex}
\ifPDFTeX
  \usepackage[T1]{fontenc}
  \usepackage[utf8]{inputenc}
  \usepackage{textcomp} 
\else 
  \usepackage{unicode-math}
  \defaultfontfeatures{Scale=MatchLowercase}
  \defaultfontfeatures[\rmfamily]{Ligatures=TeX,Scale=1}
\fi
\usepackage{lmodern}
\ifPDFTeX\else  
\fi
\IfFileExists{upquote.sty}{\usepackage{upquote}}{}
\IfFileExists{microtype.sty}{
  \usepackage[]{microtype}
  \UseMicrotypeSet[protrusion]{basicmath} 
}{}
\makeatletter
\@ifundefined{KOMAClassName}{
  \IfFileExists{parskip.sty}{%
    \usepackage{parskip}
  }{
    \setlength{\parindent}{0pt}
    \setlength{\parskip}{6pt plus 2pt minus 1pt}}
}{
  \KOMAoptions{parskip=half}}
\makeatother
\usepackage{xcolor}
\makeatletter
\ifx\paragraph\undefined\else
  \let\oldparagraph\paragraph
  \renewcommand{\paragraph}{
    \@ifstar
      \xxxParagraphStar
      \xxxParagraphNoStar
  }
  \newcommand{\xxxParagraphStar}[1]{\oldparagraph*{#1}\mbox{}}
  \newcommand{\xxxParagraphNoStar}[1]{\oldparagraph{#1}\mbox{}}
\fi
\ifx\subparagraph\undefined\else
  \let\oldsubparagraph\subparagraph
  \renewcommand{\subparagraph}{
    \@ifstar
      \xxxSubParagraphStar
      \xxxSubParagraphNoStar
  }
  \newcommand{\xxxSubParagraphStar}[1]{\oldsubparagraph*{#1}\mbox{}}
  \newcommand{\xxxSubParagraphNoStar}[1]{\oldsubparagraph{#1}\mbox{}}
\fi
\makeatother

\usepackage{longtable,booktabs,array}
\usepackage{calc} 
\usepackage{etoolbox}
\makeatletter
\patchcmd\longtable{\par}{\if@noskipsec\mbox{}\fi\par}{}{}
\makeatother
\IfFileExists{footnotehyper.sty}{\usepackage{footnotehyper}}{\usepackage{footnote}}
\makesavenoteenv{longtable}
\usepackage{graphicx}
\makeatletter
\def\maxwidth{\ifdim\Gin@nat@width>\linewidth\linewidth\else\Gin@nat@width\fi}
\def\maxheight{\ifdim\Gin@nat@height>\textheight\textheight\else\Gin@nat@height\fi}
\makeatother
\setkeys{Gin}{width=\maxwidth,height=\maxheight,keepaspectratio}
\makeatletter
\def\fps@figure{htbp}
\makeatother

\makeatletter
\@ifpackageloaded{caption}{}{\usepackage{caption}}
\AtBeginDocument{%
\ifdefined\contentsname
  \renewcommand*\contentsname{Table of contents}
\else
  \newcommand\contentsname{Table of contents}
\fi
\ifdefined\listfigurename
  \renewcommand*\listfigurename{List of Figures}
\else
  \newcommand\listfigurename{List of Figures}
\fi
\ifdefined\listtablename
  \renewcommand*\listtablename{List of Tables}
\else
  \newcommand\listtablename{List of Tables}
\fi
\ifdefined\figurename
  \renewcommand*\figurename{Figure}
\else
  \newcommand\figurename{Figure}
\fi
\ifdefined\tablename
  \renewcommand*\tablename{Table}
\else
  \newcommand\tablename{Table}
\fi
}
\@ifpackageloaded{float}{}{\usepackage{float}}
\floatstyle{ruled}
\@ifundefined{c@chapter}{\newfloat{codelisting}{h}{lop}}{\newfloat{codelisting}{h}{lop}[chapter]}
\floatname{codelisting}{Listing}

\makeatother
\makeatletter
\@ifpackageloaded{caption}{}{\usepackage{caption}}
\@ifpackageloaded{subcaption}{}{\usepackage{subcaption}}
\makeatother

\ifLuaTeX
  \usepackage{selnolig}  
\fi
\usepackage[]{natbib}
\usepackage{bookmark}

\IfFileExists{xurl.sty}{\usepackage{xurl}}{} 
\hypersetup{
  pdftitle={Empirical-Bayes spectral partial pooling across related tasks},
  pdfauthor={Lorenzo Mauri},
  pdfkeywords={Empirical Bayes, High-dimensional, Latent variable model, Multi-task, Singular value decomposition},
  colorlinks=true,
  linkcolor={blue},
  filecolor={Maroon},
  citecolor={Blue},
  urlcolor={Blue},
  pdfcreator={LaTeX via pandoc}}

\newcommand{\anon}{1}

\usepackage{amsmath,amssymb}
\usepackage{algorithm}
\usepackage{algpseudocode}

\usepackage{dirtytalk}

\algrenewcommand{\algorithmicrequire}{\textbf{Input:}}

\usepackage{comment}

\begin{document}

\def\spacingset#1{\renewcommand{\baselinestretch}%
{#1}\small\normalsize} \spacingset{1}


\if1\anon
{
  \title{\bf Empirical-Bayes spectral partial pooling across related tasks}
  \author{Lorenzo Mauri\thanks{
    This research was partially supported by the Office of Naval Research (Grant N00014-24-1-2626), and by the European Research Council under the European Union’s Horizon 2020 research and innovation programme (grant agreement No 856506). The author is grateful to David B.~Dunson for useful discussions.}\hspace{.2cm}\\
    Department of Statistical Science, Duke University}
  \maketitle
} \fi

\if0\anon
{
  \bigskip
  \bigskip
  \bigskip
  \begin{center}
    {\LARGE\bf Empirical-Bayes spectral partial pooling across related tasks}
\end{center}
  \medskip
} \fi

\bigskip
\begin{abstract}
Spectral methods are central to high-dimensional statistics and machine learning,
underlying procedures for covariance estimation, matrix denoising, representation
learning, clustering, and latent variable modeling. In this work, we focus primarily
on spectral estimators for high-dimensional factor models, where leading singular
vectors of the data matrix are used to estimate latent structure and covariance
parameters. In many modern applications, however, data are collected across related
but heterogeneous tasks, studies, domains, or populations. Applying such estimators
separately to each task can lead to unstable estimates when sample sizes are limited
relative to dimension, while completely pooling the data can obscure meaningful
task-specific structure.
We introduce Hierarchical Spectral Shrinkage (\texttt{HSS}), a scalable empirical-Bayes
framework for partially pooling task-specific spectral estimators. The method
regularizes the leading spectral directions of each task toward a data-adaptively
learned common basis, while allowing the amount of shrinkage to vary across tasks
and spectral components. The proposed estimator arises as the posterior mean in a
surrogate Bayesian regression formulation of the empirical singular vectors. 
For factor models, \texttt{HSS} yields partially pooled estimators of the group-specific
loading spaces and covariance matrices and can be combined with different spectral
estimators by replacing their empirical singular vectors with hierarchically
regularized counterparts. More broadly, the same construction provides a mechanism
for extending spectral estimators to collections of related but heterogeneous
datasets. We demonstrate substantial improvements in estimation accuracy and
out-of-sample performance over separate, fully pooled, and shared-subspace approaches
in synthetic  experiments and multi-study gene-expression data.

\end{abstract}

\noindent%
\vfill

\newpage
\spacingset{1.8} 

\section{Introduction}

In many modern applications, the same variables are measured across related studies, populations, or experimental conditions. Their covariance matrices may exhibit similar dominant patterns while retaining meaningful differences. Estimating each covariance matrix separately can be unstable when the number of variables is large relative to the study sample size, whereas pooling all observations can obscure these differences. Similarities in covariance structure can often be described through similarities in principal axes \citep{hoff_09}. Our goal is to exploit these similarities to stabilize group-specific covariance estimates while preserving differences in their dominant patterns of variation.
We focus on factor models, which represent each group-specific covariance as a low-rank signal component plus residual variation.

For each group $g=1,\dots,G$, let $Y_g\in\mathbb R^{n_g\times p}$ denote the observed data matrix. We represent it as
\begin{equation}
    Y_g=F_g\Lambda_g^\top+E_g,
\end{equation}
where $F_g \in \mathbb{R}^{n_g\times k_g}$ contains the latent factors, $\Lambda_g \in \mathbb{R}^{p\times k_g}$ contains the factor loadings, and $E_g \in \mathbb{R}^{n_g\times p}$ contains the residual errors, with $k_g \ll p$. Assuming that the factors have an identity covariance matrix and are uncorrelated with the residual errors, the group-specific covariance is
\[
\Sigma_g = \Lambda_g\Lambda_g^\top + \Psi_g,
\]
where $\Psi_g$ is diagonal and contains the residual error variances.

A broad class of factor-model estimators obtains factors and loadings from the singular value decomposition (SVD) of the observed data matrix.
Let
$A_g$ be the diagonal matrix of its leading $k_g$ singular values,
and let $U_g$ and $V_g$ contain the corresponding left and right
singular vectors. The principal component (\texttt{PC}, \citet{bai_03}) estimates of the factors and loadings in
group $g$ are
\[
\widehat F_{PC,g}=\sqrt{n_g}\,U_g,
\qquad
\widehat\Lambda_{PC,g}=\frac{V_gA_g}{\sqrt{n_g}}.
\]
Several extensions regularize these spectral estimates. \citet{Barigozzi2018ConsistentEO} regularize principal component estimates when the latent dimension is overestimated. \citet{fable} use \texttt{PC} estimates of the factors and infer the loadings through surrogate Bayesian regressions, yielding regularized loading estimates. \citet{eigenbayes} extend this construction to overfitted latent dimensions with adaptive regularization across components. 

Related approaches incorporate auxiliary information \citep{basil}, accommodate multiple modalities \citep{fama}, or target binary data \citep{flair}. Applying these methods separately to each group, however, does not exploit similarities across groups.

Methods for grouped data impose different structures on how information is shared. The common principal components model assumes common eigenvectors across groups while allowing group-specific eigenvalues \citep{flury84}. Extensions allow selected eigenvectors or eigenspaces to be shared across some or all groups \citep{flury_87, schott_91, schott_97, boik02}. Multi-study factor models \citep{msfa, bmsfa, vi_bmsfa, sufa, combinatorial_msfa, adaptive_fa, blast} and related matrix decomposition methods \citep{jive, ajive} decompose systematic variation into shared and study-specific components. 

\citet{franks_hoff_2019} estimate a common low-dimensional subspace containing the group-specific signal variation under a spiked covariance model. Their approach permits different group-specific eigenvectors and eigenvalues within that subspace. However, it forces the signal directions to remain within the common subspace. Shrinkage offers an alternative way to borrow information by encouraging similarity while allowing departures from the shared representation. 

\citet{hoff_09} allows approximate sharing by placing a hierarchical distribution on group-specific eigenvector matrices and shrinking them toward a common mode. \citet{hoff_09}'s method relies on Gibbs sampling for posterior computation, which requires approximations due to the intractability of some full conditional distributions and may become inefficient when $p$ is large. Moreover, full rank covariance estimation is unstable for high dimensional scenarios.

We introduce Hierarchical Spectral Shrinkage (\texttt{HSS}), an empirical Bayes framework for partially pooling the spectral representations, and employ it to extend spectral estimators for factor models to grouped data. We reformulate the problem of learning the principal directions of variation as a surrogate regression task. Next, we place a hierarchical prior distribution on the group-specific principal directions sharing a common mean. The common mean and group-shrinkage parameters are learnt via an empirical Bayes procedure. The final estimate is a linear combination of the empirical singular vectors and the mean basis, which can be interpreted as the posterior mean in the surrogate Bayesian regression model. In particular, for each task \(g\), \texttt{HSS} 
replaces the empirical spectral directions \(V_g\) by a regularized estimate
\begin{equation}\label{eq:V_g_hat}
\widehat{V}_g
=
V_g S_g
+
V_0 R_g (I_{k_g}-S_g),    
\end{equation}
where $V_0$ is a common spectral representation estimated using all groups, $R_g$ aligns that representation with the group-specific directions, and the diagonal matrix $S_g$ controls shrinkage separately for each group and spectral component. When $S_g=I_{k_g}$, the estimator retains the empirical representation $V_g$; when $S_g=0$, it shrinks completely toward the aligned common representation $V_0R_g$. We construct group-specific loading and covariance estimates by substituting $V_g$ with $\widehat V_g$  in a spectral factor-model estimator.
Differently from shared-component decompositions, \texttt{HSS} shrinks each group's parameters towards a common mean instead of decomposing variation into common and group-specific loading components. We demonstrate the usefulness of our method in a variety of synthetic experiments and on a multi-study gene-expression dataset.

Other spectral procedures provide potential applications of this construction.  Many covariance and matrix-denoising estimators retain the empirical eigenvectors while regularizing the corresponding eigenvalues of the empirical covariance matrix \citep{ledoit_wolf_04, ledoit_wolf_12, ledoit_wolf_17, ledoit_wolf_20, ledoit_wolf_kiss, donoho2018optimal, chatterjee2015matrix, gavish2017optimal}.
These methods suggest further applications of spectral partial pooling across related datasets. Such extensions would require adapting the surrogate regression formulation and hyperparameter estimation to these problems.

\section{Method}


\subsection{General methodology}\label{subsec:general_methodology}

We assume the $i$-th observation from the $g$-th group is generated as follows
\begin{equation} \label{eq:fact_model}        y_{g, i} = \Lambda_g f_{g, i} + \epsilon_{g,i}, 
\quad \epsilon_{g,i}  \sim N_p\left(0, \Psi_g \right),\quad (i=1, \dots, n_g; g=1, \dots, G). \end{equation}
where, $g$ indexes the group, $n_g$ is sample size, $k_g$ is latent dimension, $\Lambda_g \in \mathbb R^{p \times k_g}$ is the loadings matrix, and $ \Psi_g = \text{diag}(\sigma_{g, 1}^2, \dots, \sigma_{g, p}^2)$ is the matrix of idiosyncratic variances of group $g$ respectively. Further, we assume 
\begin{equation}
    E[f_{g, i}] = 0, \quad cov(f_{g, i}) = I_{k_g}, \quad (i=1, \dots, n_g; g=1, \dots, G).
\end{equation}
Therefore, integrating out the latent factors, we have 
\begin{equation}
    E[y_{g, i}] = 0, \quad cov(y_{g, i}) = \Lambda_g \Lambda_g^\top + \Psi_g, \quad (i=1, \dots, n_g; g=1, \dots, G).
\end{equation}
First, let the $g$-th group data matrix 
$$
Y_g = \begin{bmatrix}
    y_{g,1} & \cdots & y_{g, n_g}
\end{bmatrix}^\top \in \mathbb R^{n_g \times p}
$$
and consider its compact singular value decomposition $Y_g = U_g A_g V_g^\top + U_{g \perp} A_{g \perp} V_{g \perp}^\top$, where $A_g = \text{diag}\big(a_{g,1}, \cdots, a_{g, k_{g}}\big), A_{g \perp} = \text{diag}\big(a_{g,k_g+1}, \cdots, a_{g, \min\{n_g, p\} }\big)$ are diagonal matrices with the leading $k_g$ singular values of $Y_g$ and the remaining $\min\{n_g, p\} - k_g$ ones respectively, while $U_g, V_g$ and $U_{g \perp}$ and $V_{g \perp}$ are the matrices of the corresponding left and right singular vectors. Note that assuming the data has been centered, the empirical covariance of $Y_g$ is given by 
$\hat \Sigma_g = V_g D_g V_g^\top +  V_{g \perp} D_{g \perp} V_{g \perp}^\top$, where $D_g = A_g^2 / n_g$ and $D_{g \perp} = A_{g \perp}^2 / n_g$.

As discussed in the introduction, various estimators of $\Lambda_g$ (and other unknown quantities) are functions of $U_g, A_g$, and $V_g$. For instance, the PC-estimator for $\Lambda_g$ is 
\begin{equation}
    \widehat{\Lambda}_{PC, g} = \frac{V_g A_g}{\sqrt{n_g}}.  
\end{equation}
Instead, we suggest the \texttt{HSS} estimator
\begin{equation}
  \widehat{\Lambda}_{\texttt{HSS}, g} =  \frac{\widehat V_{g} A_g}{\sqrt{n_g}}.  
\end{equation}
where $V_g$ is replaced with 
\begin{equation}\label{eq:V_g_hat_2}
   \widehat{V}_{g} =    V_{g} S_g +  V_0 R_g (I_{k_g} - S_g).
    \end{equation}
   Equation \eqref{eq:V_g_hat_2} defines $  \widehat{V}_{g}$ to be a linear combination of the leading $k_g$ right singular vectors $V_{g}$ and a rotation of the mean basis $V_0$. $S_g$ is a  diagonal matrix where the $l$-th element encodes the shrinkage of the $l$-th singular vector towards a common basis.  Next, we motivate the choice of \eqref{eq:V_g_hat_2} and propose a data adaptive approach to estimate $S_g$, $V_0$, and $R_g$. We note that for simplicity, we apply the \texttt{HSS} methodology to the PC-estimators but it could be applied to any estimator for factor models based on the singular value decomposition of the data matrix, e.g. the ones presented in \citet{Barigozzi2018ConsistentEO, fable, eigenbayes, basil}.

\subsection{Surrogate regression formulation}\label{subsec:surrogate_regression}
This section presents a regression formulation for right singular vectors and its hierarchical extension to group setting that motivates the form of the estimator in \eqref{eq:V_g_hat_2}. 
 Let $X_g = U_{g} A_{g}$, then \begin{equation*}
     V_{g} =
 \underset{\tilde V_g \in \mathbb R^{p \times k_g}}{\operatorname{argmin}}     ||Y_g - X_g \widetilde V_g^\top||_F^2,
\end{equation*} 
where $||\cdot||_F$ denotes the Frobenius norm. 
Hence, $ V_{g}$ can be interpreted as the posterior mean of the regression coefficients matrix 
in the surrogate linear regression model 
\begin{equation}\label{eq:surrogate_regression}
    Y_g \sim N_{n_g, p} \big(X_g \widetilde V_g^\top, I_{n_g} , \Psi_g\big).
\end{equation}
under the prior 
\begin{equation*}\label{eq:hier_prior_V_g}
       \widetilde V_g \mid \Psi_g, \phi^2 
       \sim   N_{p, k_g}  \big(\mathbf{0}_{p \times k_g},  \Psi_g,  \phi^2 I_{k_g}\big),
\end{equation*}
in the limiting case $\phi^2 \to \infty$.
With this intuition in mind, we develop an estimator that shares information across groups. In particular, with grouped-data, it is natural to adopt a hierarchical prior on the $V_g$'s, borrowing information across groups:
\begin{equation}\label{eq:hier_prior_V_g}
       \widetilde V_g \mid V_0, R_g, \Psi_g, \Phi_g \sim   N_{p, k_g}  \big(V_0 R_g,  \Psi_g,  \Phi_g\big),
\quad    (g = 1, \dots, G),
\end{equation}
where $\Phi_g = \text{diag}\big(\phi_{g, 1}^2, \cdots, \phi_{g, k_g}^2\big)$ controls the prior shrinkage across each latent dimension, $V_0 \in \mathbb R^{p \times k_0}$ is the \say{common basis} matrix. The matrix $R_g \in \mathbb R^{k_0 \times k_g}$ aligns the common basis $V_0$ with the axis of variation of the data in group $g$. For identifiability purposes, we constrain $R_g$ to be a semi-orthogonal matrix if $k_0 \neq k_g$ or an orthogonal matrix when $k_0 = k_g$. 
A straightforward point estimator for $ V_g $ is given by the posterior mean under the surrogate model:
\begin{equation}\label{eq:V_hat_g}
    \widehat V_g = E[\widetilde V_g\mid -] = V_{g}S_g +  V_0 R_g (I_{k_g} - S_g),
\end{equation}
where $S_g = \text{diag}(s_{g, 1}, \dots, s_{g, k_g})$ with $s_{g,  l} = a_{g, l}^2 / (a_{g, l}^2 + 1/\phi_{g, l}^2)$. 
The simple form $\widehat V_g $ allows a clear interpretation: each main direction gets shrunk towards the common mean by a value negatively related to the corresponding empirical singular value and positively related to the prior precision $1/\phi_{g, l}$. 
We note that $ \widehat V_{g} $ is not ensured to be a semi-orthogonal matrix. One simple solution could be to replace it with its projection to the space of semi-orthogonal matrices, 
via
\begin{equation*}
     \bar V_g = 
\underset{V \in \mathcal S_{p, k_g}}{\operatorname{argmin}} || \widehat V_g  -  V||_F^2,
\end{equation*}
where \[
\mathcal S_{d_1,d_2}=\begin{cases}\left\{R\in\mathbb R^{d_1\times d_2}:R^\top R=I_{d_2}\right\}, & d_1\ge d_2,\\[2mm]\left\{R\in\mathbb R^{d_1\times d_2}:RR^\top=I_{d_1}\right\}, & d_1<d_2,\end{cases}
\]
 which is obtained by replacing the singular values of $\widehat V_{g}$ with $1$'s, in its singular value decomposition
However, since the ultimate goal is covariance estimation, we found $ \widehat V_g  \not \in \mathcal S_{p,k_g}$ not to be an issue, given that $|| \widehat V_g  -  \bar V_g ||_F$ tended to be very small in practice, and negligible differences arose between using $ \widehat V_g $ or $ \bar V_g$.
In the next subsections, we describe a data-adaptive procedure to estimate the common basis $V_0$, the aligning matrix $R_g$ and prior shrinkage parameters in $\Phi_g$.

\subsection{Estimation of the common mean and aligning matrices}\label{subsec:eb}

Next, we propose a refinement of the estimates for $V_0$ and the $R_g$'s proposed above, which uses the likelihood information stemming from \eqref{eq:surrogate_regression}. Empirical Bayes \citep{efron14} offers a natural solution to estimate hyperparameters by maximizing the log marginal likelihood. 

Let 
\begin{equation}\label{eq:marginal_likelihood}
    l\big(V_0, \big\{R_g\big\}_{g=1}^G \mid {\big\{Y_g, X_g, \Psi_g, \Phi_g\big\}_{g=1}^G}\big) = \log \prod_{g=1}^G \int p_{\text{model}}(Y_g \mid\widetilde V_g, X_g,  \Psi_g) p_{\text{prior}}(\widetilde V_g \mid V_0, R_g, \Phi_g,  \Psi_g) d \widetilde V_g
\end{equation}
be the marginal likelihood arising from surrogate regression in \eqref{eq:surrogate_regression} with the prior specification in \eqref{eq:hier_prior_V_g}, conditional on 
the residual error variances $\Psi_g$'s and the prior shrinkage hyperparameters $\Phi_g$'s,  
where $p_{\text{model}}(\cdot \mid \cdot)$ denotes the Gaussian likelihood arising from \eqref{eq:surrogate_regression} and $p_{\text{prior}}(\cdot \mid \cdot)$ denotes the Gaussian prior in \eqref{eq:hier_prior_V_g}.  
More details about \eqref{eq:marginal_likelihood} are provided in the Supplementary Material. 
Here, we consider the matrices $\Psi_g$'s and $\Phi_g$'s fixed, and the next section describes a spectral approach to estimate them. 

Thus, we estimate the hyperparameters via 
\begin{equation}\label{eq:eb_estimate}
   \big(\widehat V_0, \big\{\widehat R_g\big\}_{g=1}^G\big) =   \underset{V_0 \in \mathbb R^{p \times k_0}, \big\{R_g \in \mathcal S_{k_0, k_g}\big\}_{g=1}^G }{\operatorname{argmax}}  l\big(V_0, \big\{R_g\big\}_{g=1}^G \mid {\big\{Y_g, X_g, \Psi_g, \Phi_g\big\}_{g=1}^G}\big).
\end{equation}
We approximate the solution to \eqref{eq:eb_estimate} via an alternating scheme, that is by iterating between the following steps until convergence.
\begin{enumerate}
    \item Given the estimates $\big\{\widehat R_g\big\}_{g=1}^G$, we update the estimate for $V_0$ via
\begin{equation}\label{eq:V_0_hat}
\begin{aligned}
    &\widehat V_0 =  \big[\hat v_{0,1}, \cdots, \hat v_{0,p}\big]^{\top} =  \underset{V_0 \in \mathbb R^{p \times k_0}}{\operatorname{argmax}} \   l\big(V_0, \big\{\widehat R_g\big\}_{g=1}^G \mid {\big\{Y_g, X_g, \Psi_g, \Phi_g\big\}_{g=1}^G}\big), \quad \text{where } \\
    & \hat v_{0, j} = \bigg[\sum_{g=1}^G \bigg\{ \frac{1}{\hat \sigma_{gj}^2} \widehat  R_g \big(I_{k_g} + A_g^2 \Phi_g\big)^{-1} A_g^2 \widehat R_g^\top \bigg\} \bigg]^{-1} \bigg[\sum_{g=1}^G \bigg\{ \frac{1}{\hat \sigma_{gj}^2} \widehat  R_g \big(I_{k_g} + A_g^2 \Phi_g\big)^{-1} A_g^2 v_{g, j} \bigg\} \bigg].
\end{aligned}  
\end{equation}
    \item Given the estimate $\widehat V_0$, we update each $R_g$ by projecting to $\mathcal S_{k_0, k_g}$ the unconstrained maximizer $\widetilde R_g$, that is
\begin{equation}\label{eq:R_g_hat}
\begin{aligned}
    \widehat R_g =& \mathcal{P}_{\mathcal S_{k_0, k_g}} \big(\widetilde R_g\big),\text{ where } \\
    \widetilde R_g  =  \underset{ R_g \in \mathbb R^{k_0 \times k_g}}{\operatorname{argmax}} \  l_g \big(V_0, R_g, \Phi_g  \mid & Y_g, X_g,  \Psi_g\big) =  \big(  \widehat V_0^\top  \Psi_g^{-1} \widehat V_0 \big)^{-1} \widehat V_0^\top \Psi_g^{-1} V_g,
\end{aligned}
\end{equation}
and $l_g \big(V_0, R_g,  \Phi_g  \mid Y_g, X_g, \Psi_g\big) = \log \int p_{\text{model}}(Y_g \mid\widetilde V_g, X_g,  \Psi_g) p_{\text{prior}}(\widetilde V_g \mid V_0, R_g, \Phi_g, \Psi_g) d \widetilde V_g$. 

\end{enumerate}
The next section describes an initialization procedure for the parameters of the optimization routine.
When the optimization routine converges, we compute the final estimates $\widehat V_g$ via the formula in \eqref{eq:V_g_hat_2} and we refine the estimates for the residual error variance via the empirical squared norm of the columns of the matrices $\left\{\frac{Y_g - X_g \widehat V_g^\top}{\sqrt{n_g}}\right\}_{g=1}^G$.

\subsection{Spectral initialization procedure and estimation of hyperparameters}\label{subsec:hyperparams}%
This section describes an initialization procedure for the parameters of the optimization routine and how the hyperparameters are estimated. First, we construct the matrix 
\begin{equation*}
    \bar V = \begin{bmatrix}
        V_1 & \cdots & V_G
    \end{bmatrix} \in \mathbb R^{p \times (\sum_{g=1}^G k_g)}.
\end{equation*}
The initial estimate $\widehat V_0$ for the prior mean $V_0$ is obtained by taking the left singular vectors of $\bar V$ associated to its leading $k_0$ singular values. 
 \begin{remark}
     The procedure described above to obtain the initial estimate of $\widehat V_0$ has already appeared in the literature as part of \texttt{AJIVE}'s procedure \citep{ajive}. The main difference is that \texttt{AJIVE} is used to disentangle shared from view or group specific subspaces, while here we are interested in finding a regularization parameter to stabilize the estimate of the main axis of variation of each group that may span similar but potentially different subspaces. Moreover, \texttt{AJIVE} requires $k_0 \leq k_g$ for all $g = 1, \dots, G$, while we do not have such a requirement, and our procedure in \eqref{eq:k_0_hat_1} picks $k_0 \geq k_g$ for all $g = 1, \dots, G$. 
 \end{remark}
We obtain initial estimates for the $R_g$'s by solving the following Procrustes problem:
\begin{equation*}
    \widehat R_g = \arg_{R \in \mathcal S_{k_0, k_g}} \min|| V_g - \widehat V_0 R||_{F}^2.
\end{equation*}

Next, we describe a tuning procedure for the shrinkage hyperparameter. Note that under the hierarchical prior in \eqref{eq:hier_prior_V_g}, the squared deviation of the $l$-th column of $\widetilde V_g - V_0 R_g$ (for $l=1, \dots, k_g$) is given by 
$E[||\widetilde V_g^{(l)} - V_0 R_g^{(l)}||_F^2 \mid \Psi_g, \Phi_g] = \phi_{g, l}^2 tr(\Psi_g) = \phi_{g, l}^2 \sum_{j=1}^p \sigma_{g, j}^2$, where $\widetilde V_g^{(l)}$ and $R_g^{(l)}$ denote the $l$-th columns of $\widetilde V_g$ and $R_g$, respectively. We let $M_g^{(l)} = || V_g^{(l)} - \widehat V_0 \widehat R_g^{(l)}||_F^2$ be the squared deviation from the group-data $l$-th singular vector and the estimated prior mean, and estimate $tr(\Psi_g)$ via an empirical estimate of residual error variances by $ W_g = tr(\widehat \Psi_g)$, with
\begin{equation}\label{eq:hat_Psi_g}
\widehat \Psi_g = \text{diag}\big(\widehat \sigma_{g,1}^2, \dotsm \widehat\sigma_{g, p}^2\big), \quad \text{where} \quad \widehat{\sigma}_{g,j}^2 = \frac{Y_g^{(j)\top}(I_{n_g}-U_gU_g^\top)Y_g^{(j)}}{n_g}, \quad (j=1, \dots, p),  
\end{equation}
where $Y_g^{(j)}$ denotes the $j$-th column of $Y_g$. 
 Finally, we let the estimate for $\Phi_g$ be $\widehat \Phi_g = \text{diag} \big(\hat \phi_{g, 1}^2, \cdots, \hat \phi_{g, k_g}^2  \big)$, where the shrinkage parameter of the $l$-th component in the $g$-th group (i.e., the $l$-th diagonal element of $\Phi_g$) is 
\begin{equation}\label{eq:hat_phi_gl}
    \hat \phi_{g, l}^2 = \frac{M_g^{(l)}}{W_g},
\end{equation}
Finally, as for selection of the latent dimension, we rely on the joint information criterion (JIC) of \citet{chen_jic} for estimating the $k_g$'s, while to choose the dimension of the common mean $V_0$, we suggest two default alternative strategies. Let $\gamma_l$ be the $l$-th largest singular value of the matrix $\bar V$ and set $k_0$ to 
\begin{equation}\label{eq:k_0_hat_1}
   \hat k_0 = \max l, \quad \text{such that }\gamma_l^2 > 1- \epsilon, 
\end{equation}
or 
\begin{equation}\label{eq:k_0_hat_2}
   \hat k_0 = \max\{1, \tilde k_0\}, \quad  \text{where } \tilde k_0 =  \max l, \quad \text{such that }\gamma_l^2 > G(1- \epsilon), 
\end{equation}
where $\epsilon$ is a threshold value that we set to $0.25$ by default. The first criterion picks $k_0$ to include all the directions that are present in at least one of the groups. Indeed, if one direction is present in at least one of the groups, we expect its singular value to be at least $\approx 1$.  The second criterion instead only picks directions that are present in all the groups. Our numerical experiments suggest the former tends to work better with high similarity across groups' subspaces, while the latter achieves a better performance when groups' subspaces only display a moderate degree of similarity or share few common directions.

\subsection{Summary of the estimation procedure}\label{subsec:algorithm_summary}

\begin{algorithm}
\renewcommand{\baselinestretch}{1}
\footnotesize
\setlength{\parskip}{0pt}
\caption{Empirical-Bayes Hierarchical Spectral Shrinkage}
\label{alg:HSS}
\begin{algorithmic}[1]

\Require Centered matrices $\{Y_g\}_{g=1}^G$; ranks $\{k_g\}_{g=1}^G$ or a rank-selection rule; common dimension $k_0$ or threshold $\epsilon$; tolerance $\delta$.

\Statex \textbf{I. Initialization and hyperparameter tuning}
\For{$g=1,\ldots,G$}
    \State Estimate $k_g$ via the JIC if needed, and compute $Y_g=U_gA_gV_g^\top+U_{g\perp}A_{g\perp}V_{g\perp}^\top$ and $D_g=A_g^2/n_g$.
    \State Estimate $\widehat\Psi_g=\operatorname{diag}(\widehat\sigma_{g1}^2,\ldots,\widehat\sigma_{gp}^2)$ using \eqref{eq:hat_Psi_g}.
\EndFor
\State Set $\bar V=[V_1\;\cdots\;V_G]$; if $k_0$ is unknown, estimate it using \eqref{eq:k_0_hat_1} or \eqref{eq:k_0_hat_2}.
\State Let $\widehat V_0^{(0)}$ be the leading $k_0$ left singular vectors of $\bar V$.
\For{$g=1,\ldots,G$}
    \State Set $\widehat R_g^{(0)}=\arg\min_{R\in\mathcal S_{k_0,k_g}}\|V_g-\widehat V_0^{(0)}R\|_F^2$.
    \State For $l=1,\ldots,k_g$, set $\widehat\phi_{g,l}^2=\|V_g^{(l)}-\widehat V_0^{(0)}\widehat R_g^{(0,l)}\|_2^2/\operatorname{tr}(\widehat\Psi_g)$; then set $\widehat\Phi_g=\operatorname{diag}(\widehat\phi_{g,1}^2,\ldots,\widehat\phi_{g,k_g}^2)$.
\EndFor

\Statex \textbf{II. Marginal-likelihood maximization}
\State Initialize $t=0$, evaluate $\ell^{(0)}$, and hold $\{\widehat\Phi_g,\widehat\Psi_g\}_{g=1}^G$ fixed.
\Repeat
    \State Set $B_g=A_g^2(I_{k_g}+A_g^2\widehat\Phi_g)^{-1}$ for $g=1,\ldots,G$.
    \State For each $j$, update $\widehat v_{0,j}^{(t+1)}=[\sum_g\widehat\sigma_{gj}^{-2}\widehat R_g^{(t)}B_g\widehat R_g^{(t)\top}]^{-1}\allowbreak[\sum_g\widehat\sigma_{gj}^{-2}\widehat R_g^{(t)}B_gv_{g,j}]$.
    \For{$g=1,\ldots,G$}
        \State Set $\widetilde R_g^{(t+1)}=(\widehat V_0^{(t+1)\top}\widehat\Psi_g^{-1}\widehat V_0^{(t+1)})^{-1}\widehat V_0^{(t+1)\top}\widehat\Psi_g^{-1}V_g$ and $\widehat R_g^{(t+1)}=\mathcal P_{\mathcal S_{k_0,k_g}}(\widetilde R_g^{(t+1)})$.
    \EndFor
    \State Evaluate $\ell^{(t+1)}$ and set $\Delta^{(t+1)}=|\ell^{(t+1)}-\ell^{(t)}|/(Gp)$.
    \State Set $t\gets t+1$.
\Until{$\Delta^{(t)}<\delta$}
\State Set $\widehat V_0=\widehat V_0^{(t)}$ and $\widehat R_g=\widehat R_g^{(t)}$.

\Statex \textbf{III. Final estimator}
\For{$g=1,\ldots,G$}
    \State Set $\widehat S_g=\operatorname{diag}\{a_{g,l}^2/(a_{g,l}^2+1/\widehat\phi_{g,l}^2)\}_{l=1}^{k_g}$ and $\widehat V_g=V_g\widehat S_g+\widehat V_0\widehat R_g(I_{k_g}-\widehat S_g)$.
    \State Optionally project $\widehat V_g$ onto $\mathcal S_{p,k_g}$.
    \State Set $\widehat\Lambda_{\texttt{HSS},g}=\widehat V_gA_g/\sqrt{n_g}$.
    \State Set $\widehat\Psi_{\texttt{HSS},g}=\operatorname{diag}\{n_g^{-1}\|Y_g^{(j)}-U_gA_g\widehat V_g^{(j)\top}\|_2^2\}_{j=1}^p$ and $\widehat\Sigma_{\texttt{HSS},g}=\widehat\Lambda_{\texttt{HSS},g}\widehat\Lambda_{\texttt{HSS},g}^{\top}+\widehat\Psi_{\texttt{HSS},g}$.
\EndFor

\end{algorithmic}
\end{algorithm}

Algorithm~\ref{alg:HSS} summarizes the complete procedure. The moment estimates of the component-specific prior variances in \eqref{eq:hat_phi_gl} are used to initialize and tune the amount of pooling. These estimates are held fixed while the common basis and alignment matrices are refined by marginal-likelihood maximization.

\subsection{Computational considerations}
The main computational cost of \texttt{HSS} arises from the initial spectral decompositions and the subsequent low-dimensional coordinate-ascent updates. Let \(K=\sum_{g=1}^G k_g\). For each group, computing the SVD of \(Y_g\in\mathbb R^{n_g\times p}\) has complexity $\mathcal O\!\left(p n_g\min\{p, n_g\}\right) $, so that the total cost across groups is \(\mathcal O\{p \sum_g n_g\min(p, n_g)\}\). Since only the leading \(k_g\) components are required, this cost can be reduced using a truncated SVD to \(\mathcal O(\sum_g n_gp k_g)\). The initialization of the common basis requires the SVD costs \(\mathcal O(pK^2)\), or \(\mathcal O(pKk_0)\) when only the leading \(k_0\) singular vectors are computed. The remaining optimization operates only on matrices whose dimensions are determined by \(k_0\) and the \(k_g\)'s.
Each step of the optimization routine has a cost that is at most linear in $p$ and cubic in $k_0$. Finally, the evaluation of the marginal likelihood can be performed without constructing any \(p\times p\) matrices and the most expensive operation has complexity \(\mathcal O(pk_0K)\). 
Hence, since the latent dimensions are smaller than the sample sizes, the total computational cost is $ \mathcal O\left(p \big\{\sum_g n_g\min(p, n_g) +N_{\text{iter}}k_0^3\big\} \right)$ or, when truncated SVD are used, $ \mathcal O\left(p \big(\sum_g n_gk_g +N_{\text{iter}}k_0^3\big) \right)$, where $N_{\text{iter}}$ is the number of iterations until convergence.


\section{Numerical experiments}\label{sec:numerical_experiments}

This section compares the estimation accuracy of \texttt{HSS} to a number of competitors through synthetic experiments. We consider the PC-estimator applied separately to each dataset (\texttt{PC-no pool}), the same estimator applied after concatenating all groups (\texttt{PC-pool}), a partially pooled PC estimator that uses a common subspace while estimating group-specific loadings (\texttt{PC-partial pool}), the method in \citet{ajive} (\texttt{AJIVE}), and the shared-subspace estimator of \citet{franks_hoff_2019} (\texttt{SHARED-SUBSPACE}).

We report two variants of \texttt{HSS}, reflecting the two choices for selecting $k_0$. In \texttt{HSS-1}, $\widehat k_0$ is estimated using \eqref{eq:k_0_hat_1}, whereas \texttt{HSS-2} uses \eqref{eq:k_0_hat_2}. The estimated group-specific dimensions $\widehat k_g$ from the corresponding \texttt{HSS} fit are also used to tune the competing methods. For \texttt{PC-partial pool-1} and \texttt{SHARED-SUBSPACE-1}, the dimension of the common subspace is set to $\widehat k_0$ obtained from \eqref{eq:k_0_hat_1}. For the corresponding ``-2'' variants, we instead use
\[
\widehat k_{\mathrm{shared}}
=
\widehat k_0+\sum_{g=1}^G(\widehat k_g-\widehat k_0),
\]
where $\widehat k_0$ and $\widehat k_g$ are obtained from the \texttt{HSS-2} fit. We similarly consider two versions of \texttt{AJIVE}: \texttt{AJIVE-1} uses the group-specific dimensions from \texttt{HSS-1}, while \texttt{AJIVE-2}  sets the rank of the shared component to $\min\left\{\widehat k_0,\min_g(\widehat k_g-1)\right\}$, where $\widehat k_0$ is selected via \eqref{eq:k_0_hat_2}. Finally, for \texttt{PC-pool}, the latent dimension is selected by applying the JIC to the concatenated dataset. Additional details about the competing methods are provided in the Supplementary Material.

We assess estimation accuracy through the relative Frobenius error of the low-rank covariance component,
\[
\frac{\left\|\widehat{\Lambda}_g\widehat{\Lambda}_g^\top-\Lambda_g\Lambda_g^\top\right\|_F}{\left\|\Lambda_g\Lambda_g^\top\right\|_F},
\]
which rescales the estimation error by the magnitude of the true low-rank component and facilitates comparisons across different values of $p$.

We consider \(G=4\) groups generated as follows: 
\begin{equation*}
    y_{g,i}  =  \Lambda_g f_{g, i} + \epsilon_{g, i}, \quad f_{g, i} \sim N_{k_g}(0, I_{k_g}), \quad \epsilon_{g, i} \sim N_{p}(0, \Psi_g), \quad (i=1, \dots, n_g; g=1, \dots, G),
\end{equation*}
where $\Psi_g = \text{diag}\big(\sigma_{g,1}^2, \dots, \sigma_{g,p}^2\big)$ and $\sigma_{g, j}^2 \sim \text{Unif}(0.5, 5)$ for $g=1,\dots, G$ and $j=1, \dots p$. We set \(k_g=10\) and consider three mechanisms for generating \(\Lambda_g\). The parameter \(\alpha\in[0,1]\) controls the degree of similarity across groups.

\textbf{(a) Hierarchical eigenspace model.}
We generate $B_0\in\mathbb R^{p\times k}$ with $(B_0)_{jh}\overset{\mathrm{iid}}{\sim}N(0,\sigma_\lambda^2)$, and write its SVD as \(B_0=V_0D_0W_0^\top\). For each group, we first generate $
\widetilde V_g
=
\alpha V_0+\sqrt{1-\alpha^2}\,Z_g$ with $(Z_g)_{jh}\overset{\mathrm{iid}}{\sim}N(0,1/p),
$
and, then, obtain its projection $
V_g
$ onto the Stiefel manifold. We independently perturb the singular values as
$
D_g=D_0\operatorname{diag}(u_{g1},\ldots,u_{gk})$, with $
u_{gh}\overset{\mathrm{iid}}{\sim}\operatorname{Unif}(0.5,1.5)
$, 
and set
$$
\Lambda_g=V_gD_g.
$$
Thus, \(\alpha=1\) gives a common eigenspace, while smaller \(\alpha\) yields increasingly heterogeneous group-specific eigenspaces.

\textbf{(b) Classical hierarchical loading model.}
We generate a common loading matrix $\Lambda_0$, where 
$
(\Lambda_0)_{jh}\overset{\mathrm{iid}}{\sim}
N(0,\alpha^2\sigma_\lambda^2)
$ 
and group-specific deviations $\Delta_g$ 
$
(\Delta_g)_{jh}\overset{\mathrm{iid}}{\sim}
N\!\left(0,(1-\alpha^2)\sigma_\lambda^2\right),
$ independently across groups, and set
$$
\Lambda_g=\Lambda_0+\Delta_g.
$$
Therefore, \(\alpha=0\) gives independent loading matrices, whereas \(\alpha=1\) gives completely identical covariances.

\textbf{(c) Shared and group-specific directions.}
We again obtain \(V_0,D_0\) from the SVD of a \(p\times k\) matrix with iid \(N(0,\sigma_\lambda^2)\) entries and set $
r=\lfloor \alpha k \rfloor.
$ The first \(r\) directions $V_{\mathrm{sh}}$ corresponding to the first $r$ columns of $V_0$ 
are shared across groups. For \(0<r<k\), we generate $
Z_g\in\mathbb R^{p\times(k-r)}$ with $
(Z_g)_{jh}\overset{\mathrm{iid}}{\sim}N(0,1)
$, 
remove its component in the shared subspace, $
Z_g^\perp
=
\left(I_p-V_{\mathrm{sh}}V_{\mathrm{sh}}^\top\right)Z_g,
$
and define $
V_g=
\left[
V_{\mathrm{sh}},
\mathcal P_{\mathrm{St}}(Z_g^\perp)
\right].
$
Finally,
$
D_g=D_0\operatorname{diag}(u_{g1},\ldots,u_{gk})$ with $u_{gh}\overset{\mathrm{iid}}{\sim}\operatorname{Unif}(0.5,1.5)$ and let $$
\Lambda_g=V_gD_g.
$$
Thus, in this setting \(\alpha\) controls the proportion of directions that are shared across groups.

Hence, scenario (a) closely resembles the hierarchical structure in \eqref{eq:hier_prior_V_g} that we adopt to borrow information across groups. Scenario (b) corresponds to a more standard hierarchical setting with group-parameters being Gaussian perturbation of a \say{mean} parameter. Scenario (c) assumes shared and group-specific directions, mimicking multi-study scenarios with systematic variation and batch effects. 
In all the experiments, we set $\sigma_\lambda = 0.5$, we consider an unbalanced setting with \(n_g=(200,200,500,500)\) and two balanced settings \(n_g=200\) or \(500\) for all groups, and replicate each experiment $50$ times. Not all replicates produced results for every competing method. Missing runs were overwhelmingly concentrated in \texttt{SHARED-SUBSPACE}, particularly \texttt{SHARED-SUBSPACE-2}, due to numerical errors encountered when fitting the method; the other methods successfully returned estimates in essentially all runs. The reduction in completed runs for \texttt{SHARED-SUBSPACE} was most pronounced for $p=500$ and the smaller-sample configurations, although some missing runs were also observed for $p=2000$. We therefore report, for each method and configuration, the mean estimation error and corresponding uncertainty interval computed using all successfully completed replicates. Comparisons involving \texttt{SHARED-SUBSPACE} in configurations with substantially fewer completed runs should consequently be interpreted with some caution.

Across the three data-generating mechanisms, the two HSS variants exhibit complementary behavior. In general, \texttt{HSS-2} performs better when cross-group similarity is weak, or when only a small number of directions are shared, whereas \texttt{HSS-1} becomes preferable as $\alpha$ increases and the group-specific eigenspaces become more similar. This pattern is consistent across sample-size and dimensionality configurations.

\begin{figure}
\centering
\includegraphics[width=0.95\textwidth]{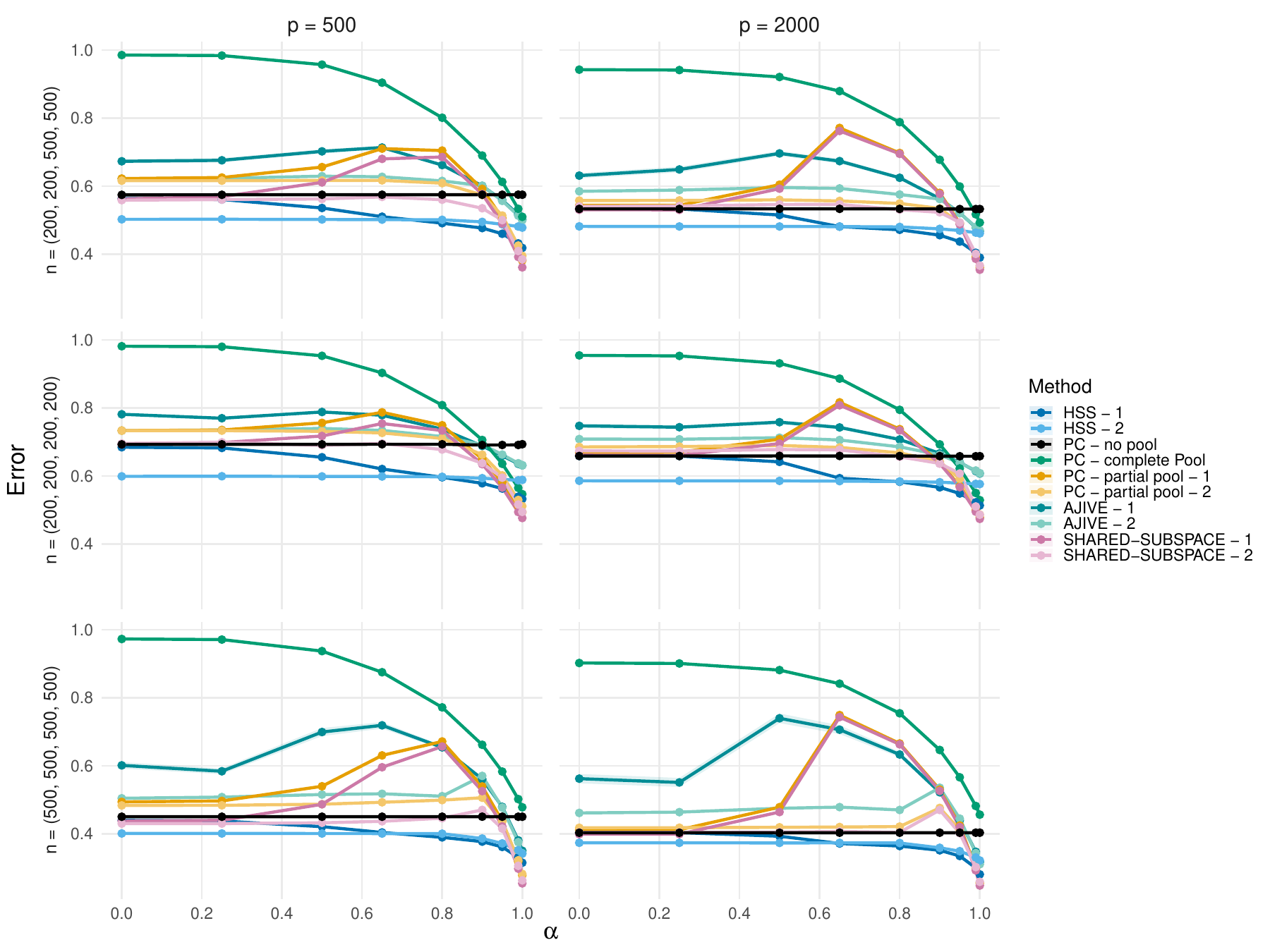}
\caption{Simulation results for the scenario (a) data-generating mechanism. Rows correspond, from top to bottom, to $n_1=n_2=200,\ n_3=n_4=500$, $n_1=n_2=n_3=n_4=200$, and $n_1=n_2=n_3=n_4=500$. Columns correspond to $p=500$ (left) and $p=2000$ (right).}
\label{fig:scenario_a}
\end{figure}

In scenario (a) (Figure \ref{fig:scenario_a}), the \texttt{HSS} estimators outperform the competing methods in most settings, except when $\alpha \geq 0.95$, where the groups are nearly identical. When groups are extremely similar to each other (i.e. $\alpha \geq 0.95$), \texttt{SHARED-SUBSPACE} and \texttt{PC-partial pool} become the best performing methods by a small margin. 

\begin{figure}
\centering
\includegraphics[width=0.95\textwidth]{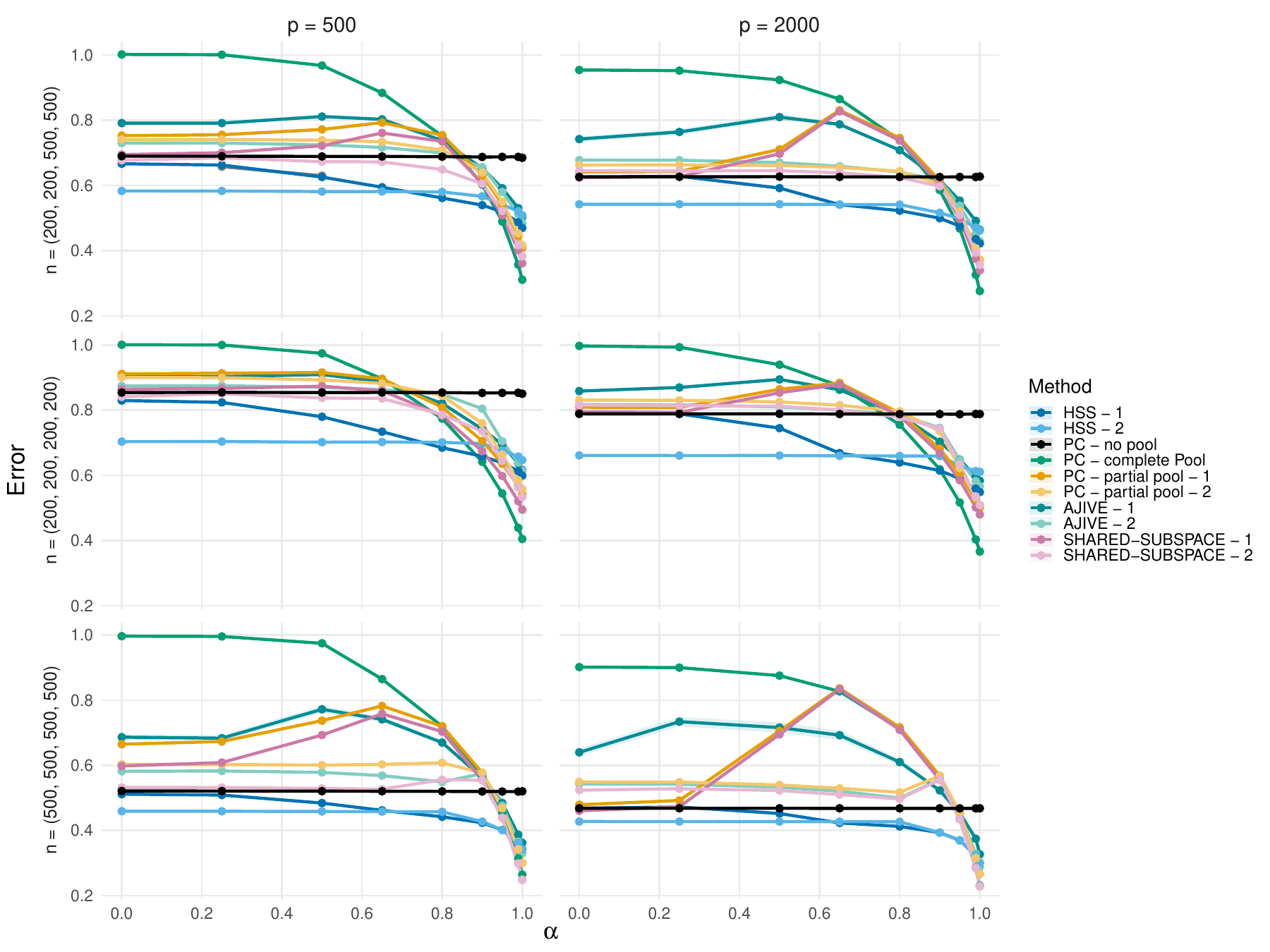}
\caption{Simulation results for the scenario (b) data-generating mechanism. Rows correspond, from top to bottom, to $n_1=n_2=200,\ n_3=n_4=500$, $n_1=n_2=n_3=n_4=200$, and $n_1=n_2=n_3=n_4=500$. Columns correspond to $p=500$ (left) and $p=2000$ (right).}
\label{fig:scenario_b}
\end{figure}

\begin{figure}
\centering
\includegraphics[width=0.95\textwidth]{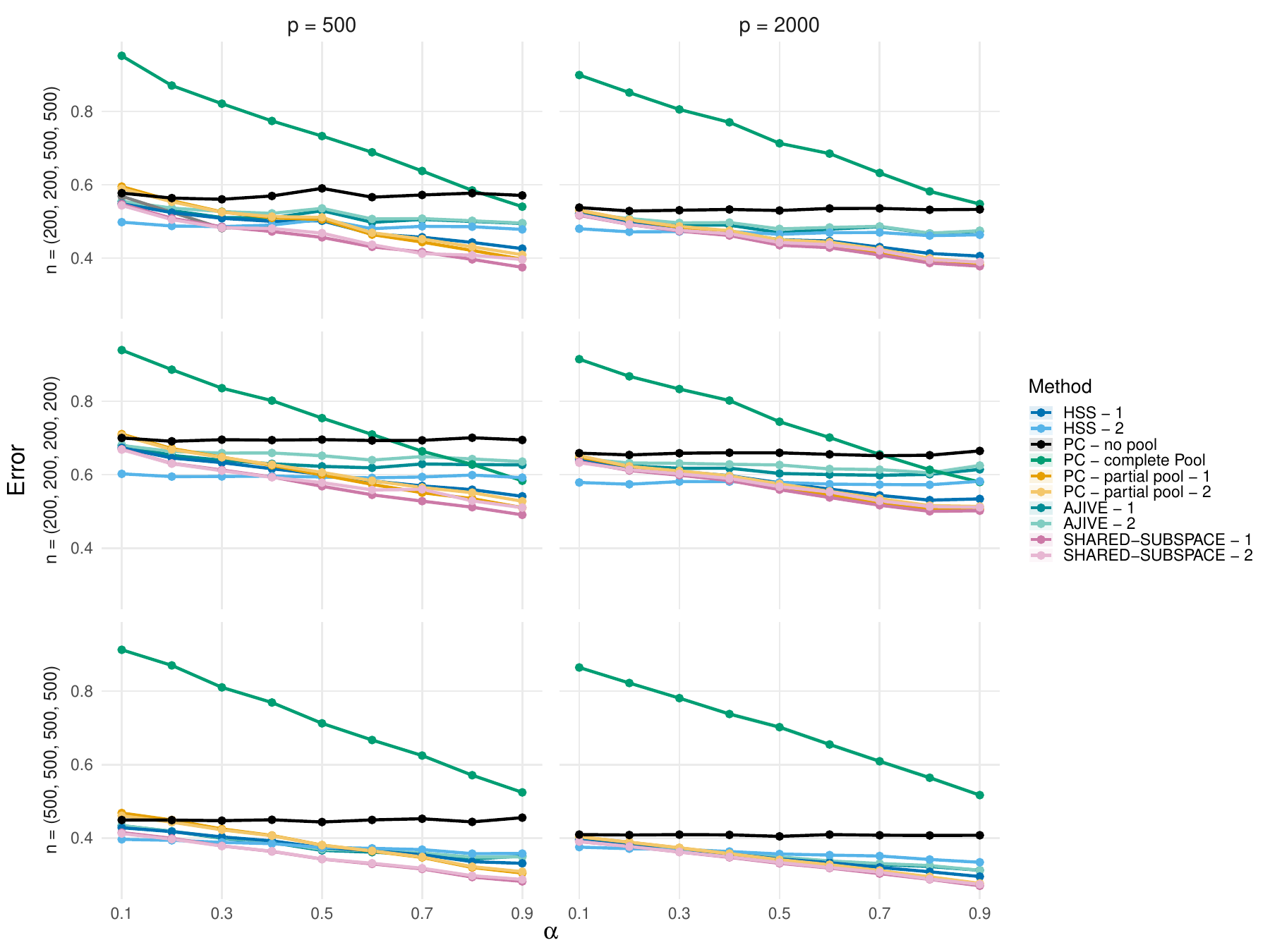}
\caption{Simulation results for the scenario (c) data-generating mechanism. Rows correspond, from top to bottom, to $n_1=n_2=200,\ n_3=n_4=500$, $n_1=n_2=n_3=n_4=200$, and $n_1=n_2=n_3=n_4=500$. Columns correspond to $p=500$ (left) and $p=2000$ (right).}
\label{fig:scenario_c}
\end{figure}

 Scenario (b) (Figure \ref{fig:scenario_b}) shows a qualitatively similar pattern. In this case, however, \texttt{Complete Pool} becomes optimal as $\alpha$ approaches one, as expected since the group-specific loading matrices, and consequently their covariance structures, become identical. Finally, in scenario (c) (Figure \ref{fig:scenario_c}), \texttt{HSS-2} performs particularly well when only a few latent directions are shared. As the number of exactly shared directions increases, methods explicitly based on a shared-subspace assumption, such as \texttt{Shared-Subspace-1} and \texttt{Shared-Subspace-2}, tend to perform best, although their advantage over \texttt{HSS-1} is generally modest.

Interestingly, \texttt{HSS} always outperforms the PCA estimator that does not borrow any information (\texttt{PC-no pool}). As expected, its advantage is larger when similarity across groups is large, but remains positive as $\alpha$ approaches $0$ and groups have completely different parameters. Similarly, \texttt{HSS} outperforms the PCA estimator that completely pools the datasets (\texttt{PC-complete pool}), except in scenario (b) for large values of $\alpha$ when the group-covariances become almost identical. 

Overall, the numerical experiments provide evidence of our methodology to successfully borrow information across groups, improving estimation accuracy, even when the similarity across groups substantially differs from the assumed hierarchical prior.

\section{Application}\label{seq:application}
We analyze three gene expression datasets derived from immune cell profiling studies. Estimating gene–gene dependencies plays a central role in advancing targeted cancer therapeutics \citep{tan}. Two of the datasets originate from the ImmGen consortium \citep{yoshida}. The first is the bulk RNA-seq dataset GSE109125, comprising 103 immune cell populations \citep{yoshida}, while the second is the microarray dataset GSE37448 \citep{elpek}. The third dataset, GSE15907 \citep{painter, desch}, is also based on microarray technology and captures expression measurements across multiple ex vivo immune lineages, primarily from adult C57BL/6 (B6) male mice.
Sample sizes across the three studies are 156, 628, and 146, respectively. 

We consider multiple experiments, where we first retain the genes that belong to the group of $c$ genes with the highest variance in each study, and randomly select $p$ of them. We let $c \in \{25\%, 50\%, 75\%\}$ and $p \in \{500, 2000\}$. We pre-process the data by applying a rank-based inverse normal transformation separately to each gene within each study. In applying the JIC criterion, we upperbound the latent dimension of each study by the number of principal components explaining at least $90\%$ of the variability. 
We perform random $80\%-20\%$ train-test splits, estimate the covariance matrix on the train set, and compute the log-likelihood using the test set. 
For each group, as well as for the overall concatenated dataset, we perform one-sided paired $t$-tests comparing the out-of-sample log-likelihood of \texttt{HSS} with that of each competing method.
\begin{figure}
\centering
\begin{subfigure}{0.43\textwidth}
    \centering
    \includegraphics[width=\textwidth]{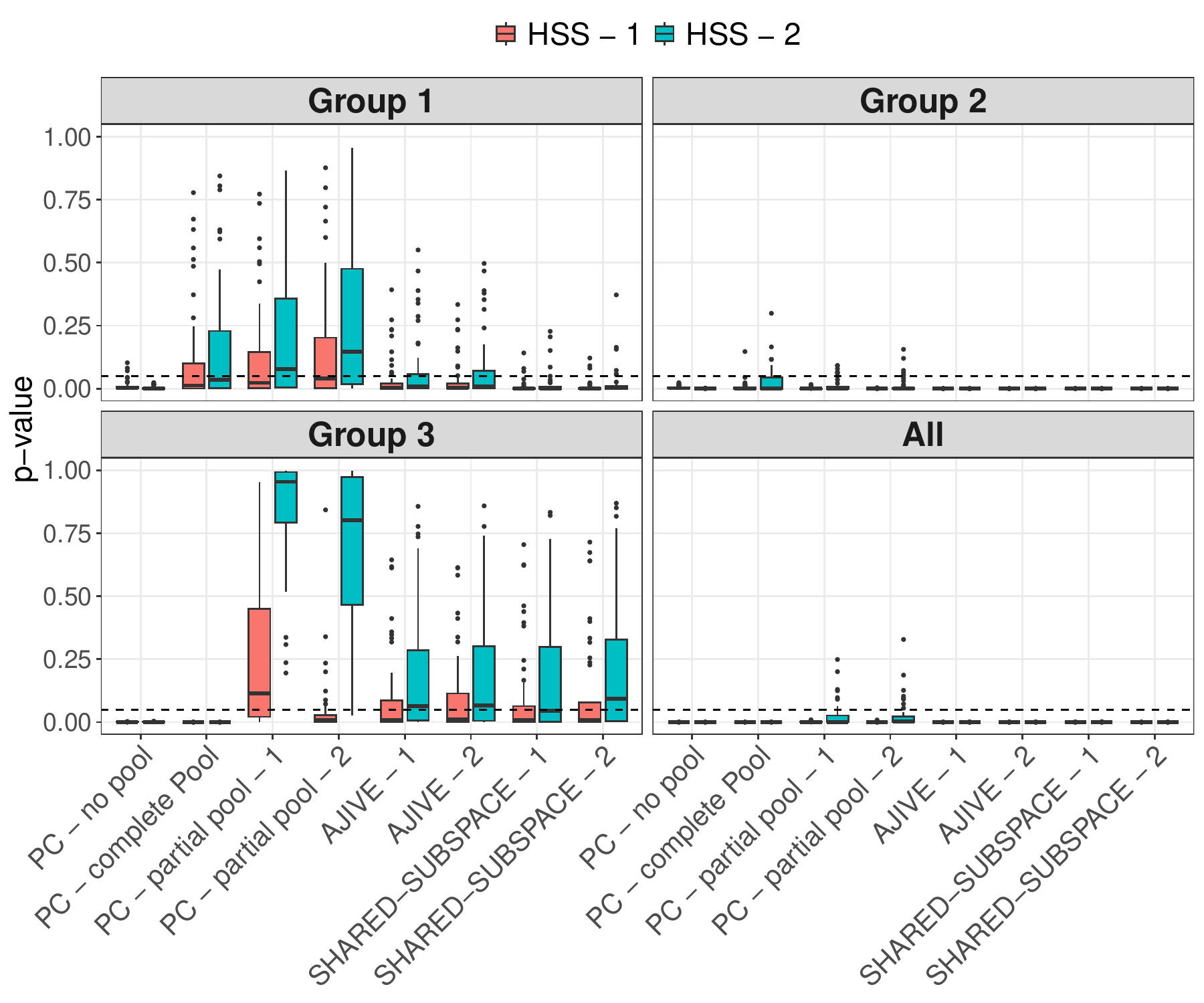}
    \caption{$c = 75\%$, $p=500$}
\end{subfigure}
\hfill
\begin{subfigure}{0.43\textwidth}
    \centering
    \includegraphics[width=\textwidth]{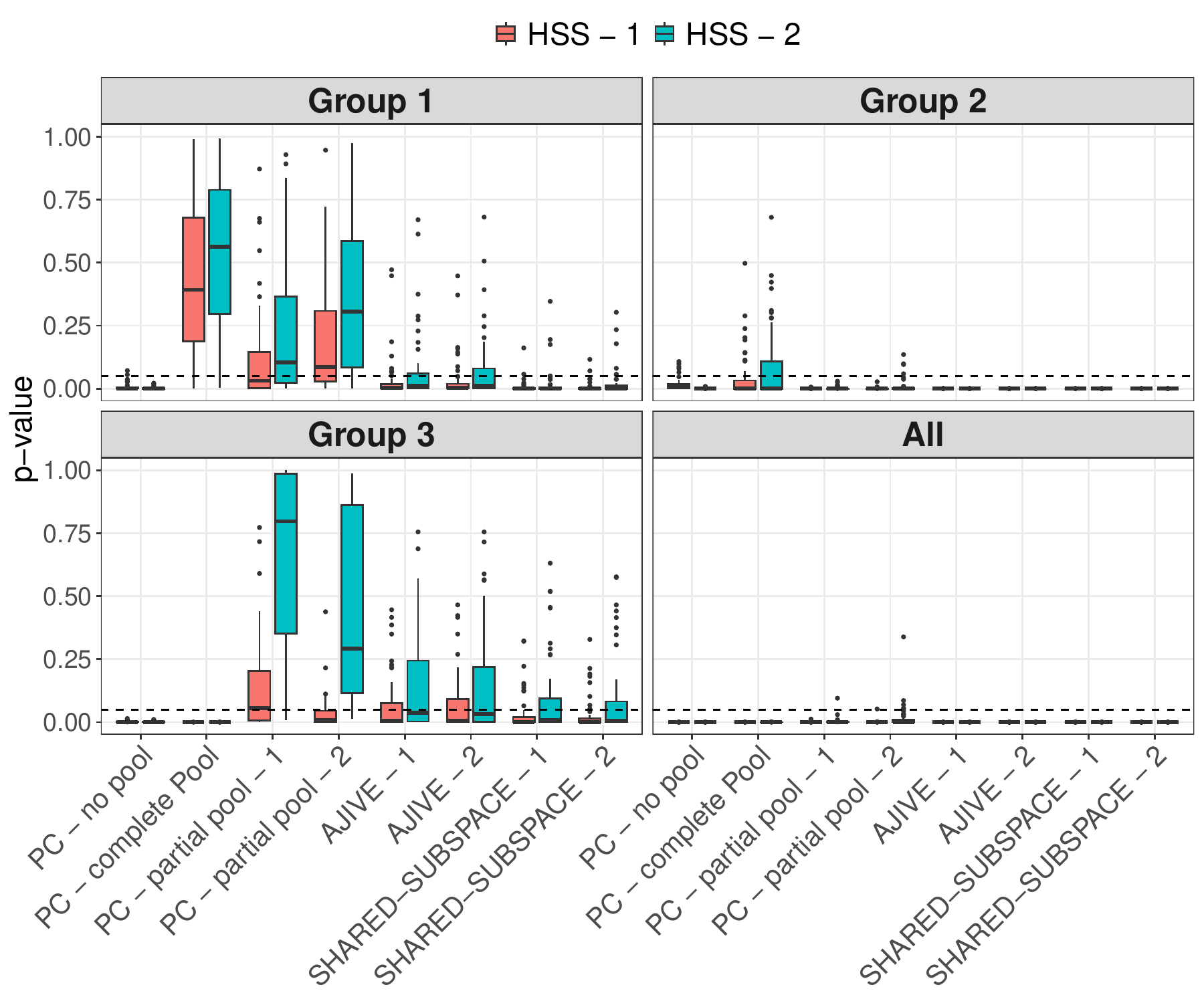}
    \caption{$c = 75\%$, $p=2000$}
\end{subfigure}

\begin{subfigure}{0.43\textwidth}
    \centering
    \includegraphics[width=\textwidth]{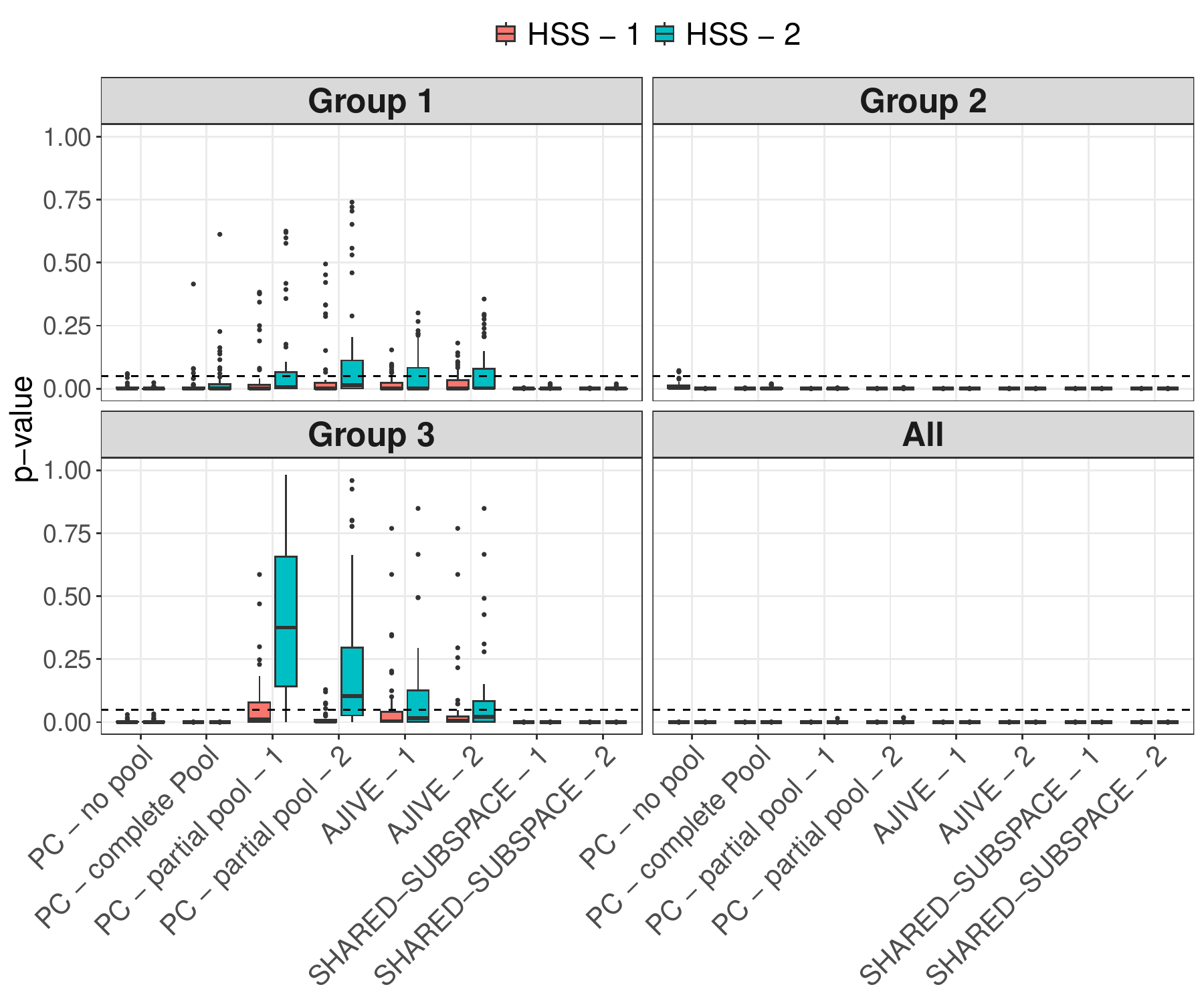}
    \caption{$c = 50\%$, $p=500$}
\end{subfigure}
\hfill
\begin{subfigure}{0.43\textwidth}
    \centering
    \includegraphics[width=\textwidth]{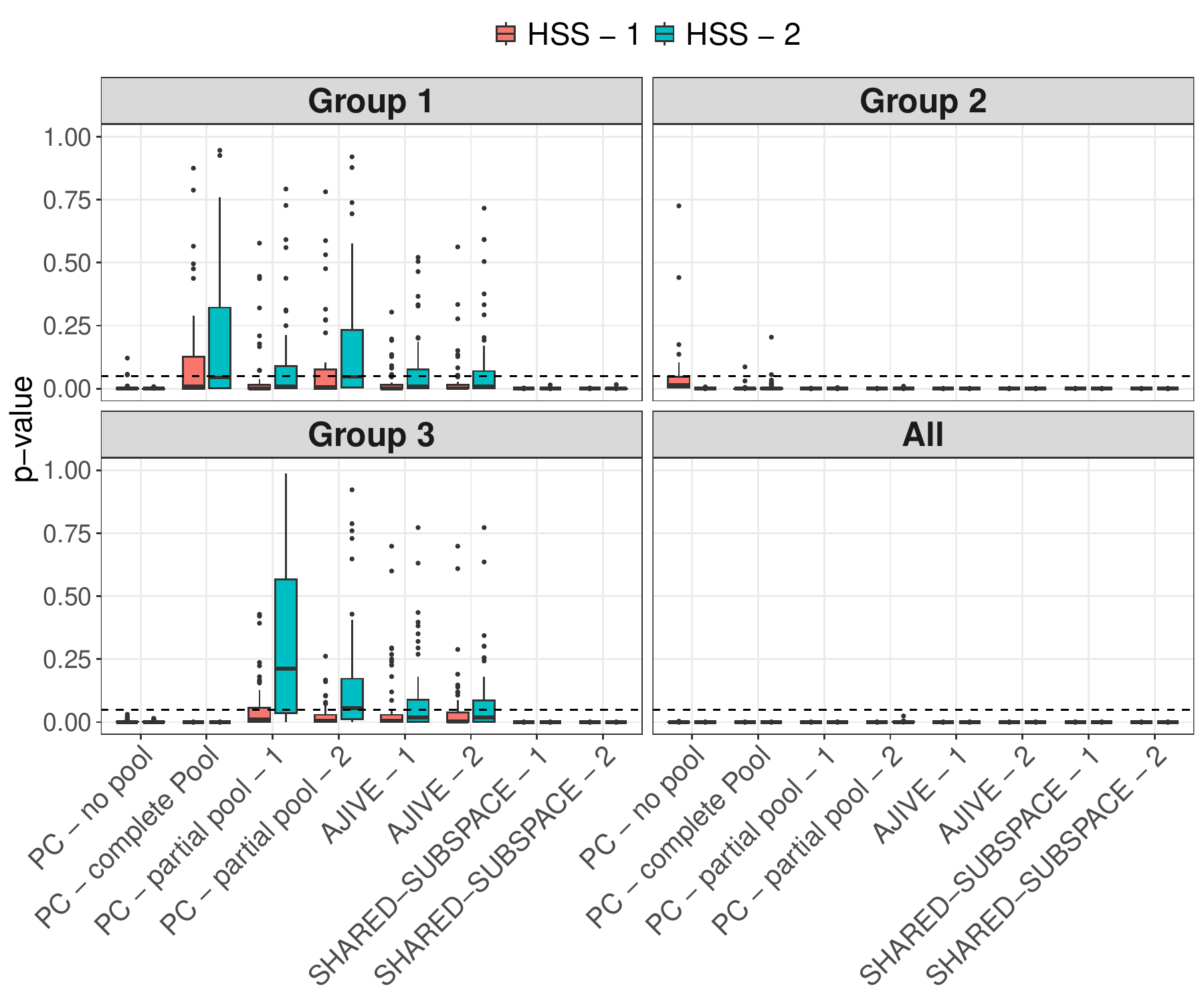}
    \caption{$c = 50\%$, $p=2000$}
\end{subfigure}

\begin{subfigure}{0.43\textwidth}
    \centering
    \includegraphics[width=\textwidth]{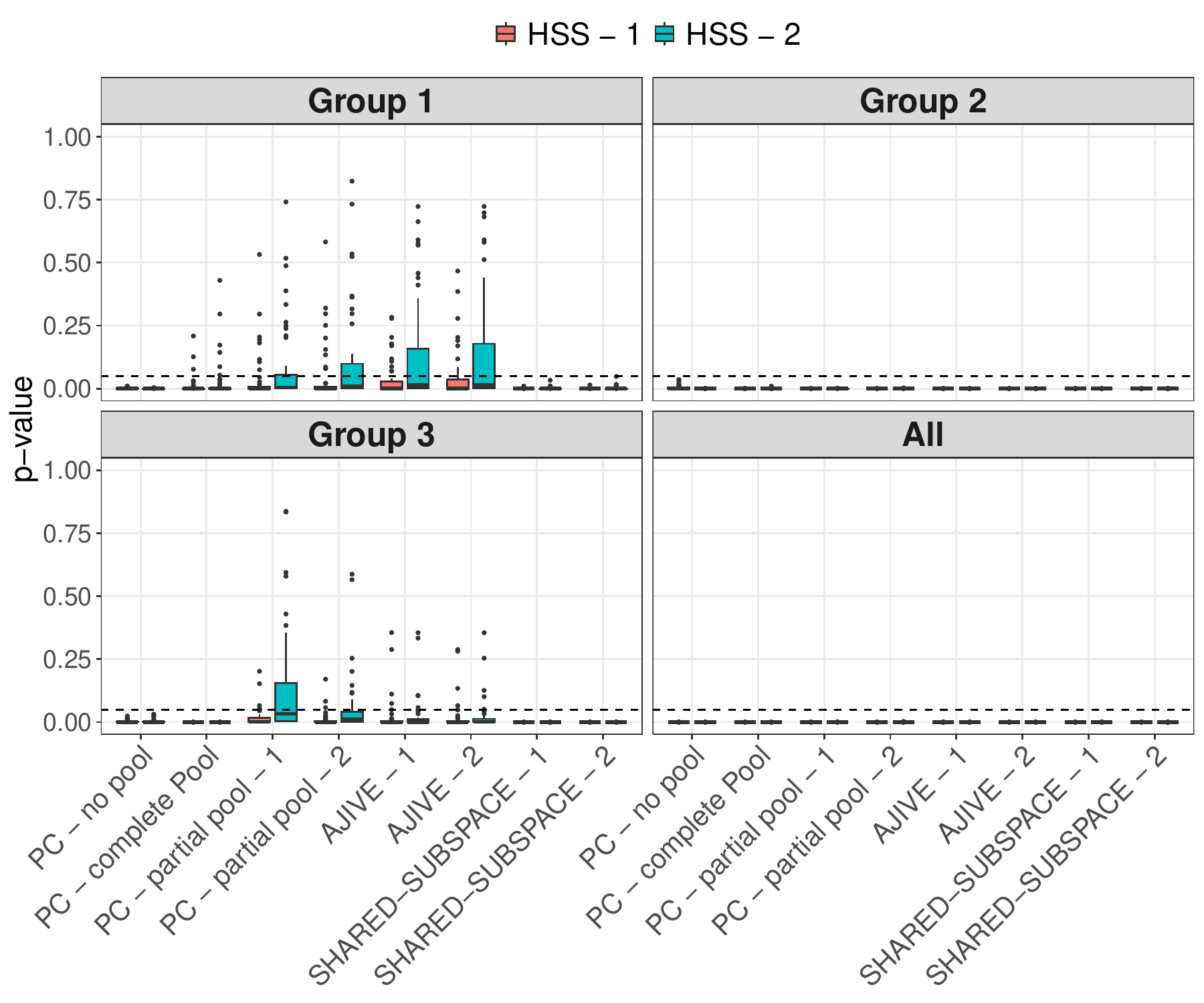}
    \caption{$c = 25\%$, $p=500$}
\end{subfigure}
\hfill
\begin{subfigure}{0.43\textwidth}
    \centering
    \includegraphics[width=\textwidth]{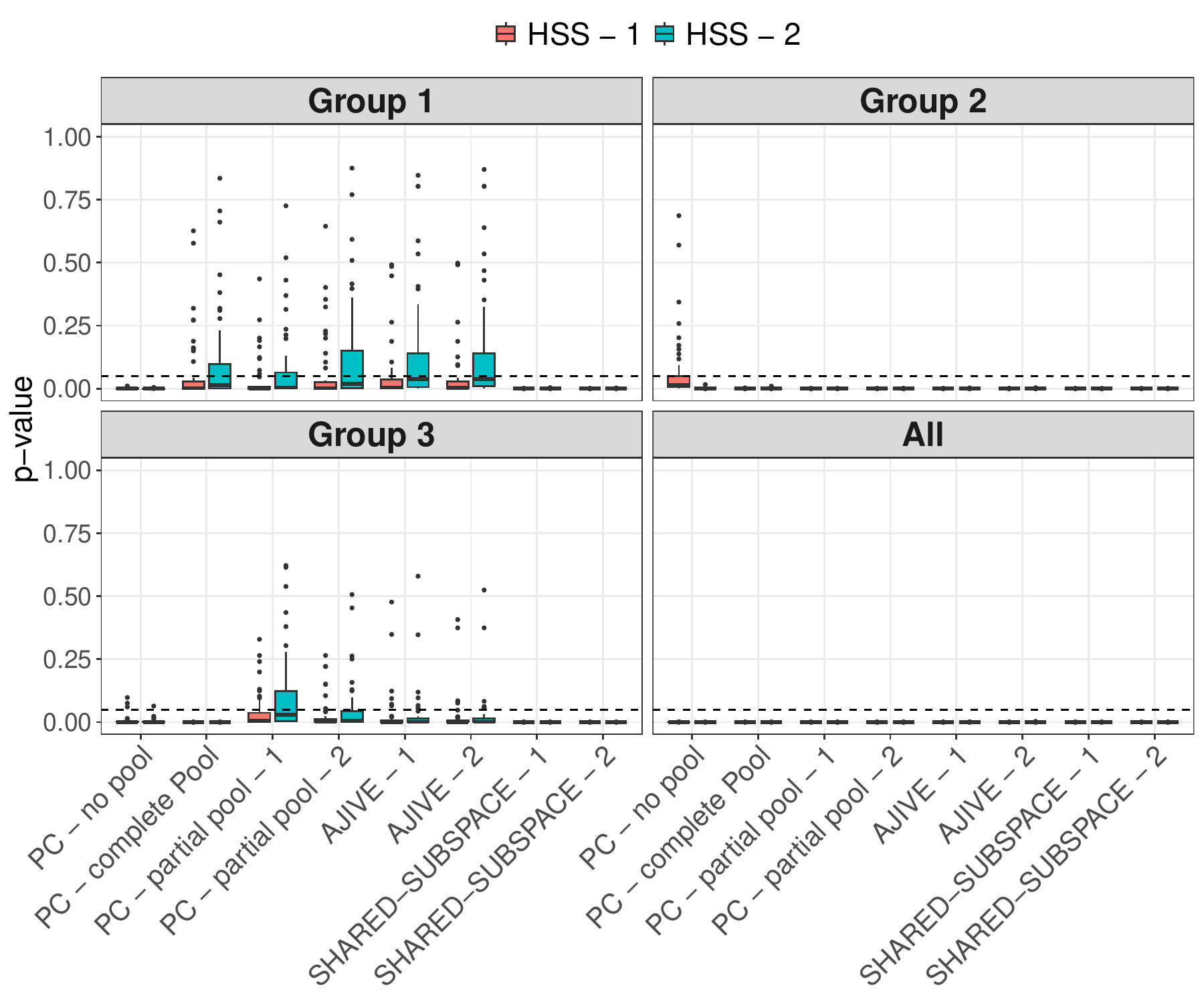}
    \caption{$c = 25\%$, $p=2000$}
\end{subfigure}

\caption{P-values of paired $t$-tests for $c$ values $75\%$, $50\%$, and $25\%$ (top to bottom) and dimensions $p=500$ (left) and $p=2000$ (right). \texttt{HSS-1} and \texttt{HSS-2} denote our methodology where $k_0$ is estimated via the criterion in \eqref{eq:k_0_hat_1} and \eqref{eq:k_0_hat_2} respectively. }\label{fig:ttest_pvalues}
\end{figure}

For each configuration, we repeat the experiments $50$ times. Results are reported in Figure \ref{fig:ttest_pvalues}. The evidence in favor of \texttt{HSS} is somewhat milder for $c=75\%$ (top row). In particular, for Groups 1 and 3, several comparisons do not provide strong evidence of an improvement, especially for \texttt{HSS}-2, which selects the shared dimension using the criterion in \eqref{eq:k_0_hat_2}. As the cutoff decreases to $c=50\%$ and $c=25\%$, the evidence becomes substantially stronger: the $p$-values concentrate below the $0.05$ significance threshold for nearly all competitors and across all groups, for both $p=500$ and $p=2000$. Overall, these results provide consistent evidence that \texttt{HSS} improves over the competing estimators, with the advantage becoming particularly clear under the more stringent variance-filtering choices.

\section{Discussion}\label{seq:discussion}
We have introduced novel spectral estimators to factor models for group data. The key aspect lies in regularizing estimates of the main axis of variation within each group by shrinking them toward a common mean, learnt using data from all groups. We have justified the novel estimator via a surrogate regression formulation of singular vectors and provided a data-adaptive procedure to estimate hyperparameters. 
There are several directions for future research. First, developing theoretical guarantees of the proposed estimator including finite sample and asymptotic performance is an interesting avenue. 
Secondly, as discussed in the introduction, the construction presented here is more general. Hence, another worthwhile extension would be to apply such technique to other estimators based on singular value decomposition. One could replace the singular vectors with the shrunk estimate proposed in Section \ref{subsec:surrogate_regression} in applying existing spectral estimators in presence of grouped data. Studying finite sample performance and theoretical properties of the resulting estimator or potentially adapting the surrogate regression formulation and hyperparameter tuning procedure to those tasks is another interesting avenue. In addition, here, we used the Bayesian regression formulation to derive a regularized point estimator, while one could exploit the posterior distribution from such formulation to characterize uncertainty about point estimates. Analyzing the empirical performance, as well as its frequentist properties in terms of coverage of the true parameters of such approach is an interesting extension.

\section*{Disclosure statement}\label{disclosure-statement}

The authors have the following conflicts of interest to declare.
 
  \bibliography{bibliography.bib}

\phantomsection\label{supplementary-material}

\newpage

\appendix

\begin{center}
    {\LARGE\bf Supplementary material for \say{Empirical-Bayes spectral partial pooling across related tasks}}

\end{center}

\section{Derivation of the log marginal likelihood in \eqref{eq:marginal_likelihood}}\label{appendix:marg_likelihood}

We now expand the marginal likelihood in \eqref{eq:marginal_likelihood}. For each group \(g\), the surrogate regression model and prior are
\begin{equation*}
Y_g\mid \widetilde V_g,X_g,\widehat \Psi_g \sim N_{n_g,p}\left(X_g\widetilde V_g^\top,I_{n_g},\widehat \Psi_g\right), \qquad \widetilde V_g\mid V_0,R_g,\widehat \Psi_g,\widehat \Phi_g \sim N_{p,k_g}\left(V_0R_g,\widehat \Psi_g,\widehat \Phi_g\right),
\end{equation*}
where \(X_g=U_gA_g\), \(Y_g=U_gA_gV_g^\top+U_{g\perp}A_{g\perp}V_{g\perp}^\top\), \(U_g^\top U_g=I_{k_g}\), and \(A_g=\operatorname{diag}(a_{g,1},\ldots,a_{g,k_g})\). Since groups are conditionally independent under the surrogate model, the log-marginal likelihood in \eqref{eq:marginal_likelihood} can be written as
\begin{equation*}
\ell\left(V_0,\{R_g,\widehat \Phi_g\}_{g=1}^G\mid \{Y_g,X_g,\widehat \Psi_g\}_{g=1}^G\right)=\sum_{g=1}^G \ell_g\left(V_0,R_g,\widehat \Phi_g\mid Y_g,X_g,\widehat \Psi_g\right),
\end{equation*}
where
\begin{equation*}
\ell_g\left(V_0,R_g,\widehat \Phi_g\mid Y_g,X_g,\widehat \Psi_g\right)=\log\int p_{\mathrm{model}}\left(Y_g\mid \widetilde V_g,X_g,\widehat \Psi_g\right)p_{\mathrm{prior}}\left(\widetilde V_g\mid V_0,R_g,\widehat \Phi_g,\widehat \Psi_g\right)d\widetilde V_g.
\end{equation*}
Integrating out \(\widetilde V_g\) gives
\begin{equation*}
Y_g\mid V_0,R_g,\widehat \Phi_g,\widehat \Psi_g \sim N_{n_g,p}\left(X_gR_g^\top V_0^\top, I_{n_g}+X_g\widehat \Phi_gX_g^\top,\widehat \Psi_g\right).
\end{equation*}
Therefore,
\begin{equation*}
\begin{aligned}
\ell_g\left(V_0,R_g,\widehat \Phi_g\mid Y_g,X_g,\widehat \Psi_g\right)
=&-\frac{n_gp}{2}\log(2\pi)-\frac{p}{2}\log\left|I_{n_g}+X_g\widehat \Phi_gX_g^\top\right|-\frac{n_g}{2}\log|\widehat \Psi_g| \\
&-\frac{1}{2}\operatorname{tr}\left[\widehat \Psi_g^{-1}\{Y_g-X_gR_g^\top V_0^\top\}^\top\left(I_{n_g}+X_g\widehat \Phi_gX_g^\top\right)^{-1}\{Y_g-X_gR_g^\top V_0^\top\}\right].
\end{aligned}
\end{equation*}
Since \(X_g=U_gA_g\), with \(U_g^\top U_g=I_{k_g}\), the determinant term satisfies
\begin{equation*}
\left|I_{n_g}+X_g\widehat \Phi_gX_g^\top\right|=\left|I_{n_g}+U_gA_g\widehat \Phi_gA_gU_g^\top\right|=\left|I_{k_g}+A_g\widehat \Phi_gA_g\right|=\left|I_{k_g}+A_g^2\widehat \Phi_g\right|,
\end{equation*}
where the final equality uses that \(A_g\) and \(\widehat \Phi_g\) are diagonal. Moreover,
\begin{equation*}
\left(I_{n_g}+X_g\widehat \Phi_gX_g^\top\right)^{-1}=U_g\left(I_{k_g}+A_g^2\widehat \Phi_g\right)^{-1}U_g^\top+\left(I_{n_g}-U_gU_g^\top\right).
\end{equation*}
Using the singular value decomposition of \(Y_g\), we have 
\begin{equation*}
Y_g-X_gR_g^\top V_0^\top=U_gA_g(V_g-V_0R_g)^\top+U_{g\perp}A_{g\perp}V_{g\perp}^\top.
\end{equation*}
Since \(U_g^\top U_{g\perp}=0\), the quadratic term decomposes as
\begin{equation*}
\begin{aligned}
&\{Y_g-X_gR_g^\top V_0^\top\}^\top\left(I_{n_g}+X_g\widehat \Phi_gX_g^\top\right)^{-1}\{Y_g-X_gR_g^\top V_0^\top\} \\
&\qquad =(V_g-V_0R_g)A_g\left(I_{k_g}+A_g^2\widehat \Phi_g\right)^{-1}A_g(V_g-V_0R_g)^\top+V_{g\perp}A_{g\perp}^2V_{g\perp}^\top.
\end{aligned}
\end{equation*}
Hence,
\begin{equation*}
\begin{aligned}
\ell_g\left(V_0,R_g,\widehat \Phi_g\mid Y_g,X_g,\widehat \Psi_g\right)
=&-\frac{n_gp}{2}\log(2\pi)-\frac{p}{2}\log\left|I_{k_g}+A_g^2\widehat \Phi_g\right|-\frac{n_g}{2}\log|\widehat \Psi_g| \\
&-\frac{1}{2}\operatorname{tr}\left[\widehat \Psi_g^{-1}(V_g-V_0R_g)A_g^2\left(I_{k_g}+A_g^2\widehat \Phi_g\right)^{-1}(V_g-V_0R_g)^\top\right] \\
&-\frac{1}{2}\operatorname{tr}\left[\widehat \Psi_g^{-1}V_{g\perp}A_{g\perp}^2V_{g\perp}^\top\right].
\end{aligned}
\end{equation*}
The final term does not depend on \(V_0\), \(R_g\), or \(\widehat \Phi_g\). Therefore, after replacing \(\widehat \Psi_g\) by the plug-in estimator \(\widehat\Psi_g\), the empirical Bayes objective in \eqref{eq:marginal_likelihood} is, up to an additive constant independent of \(V_0,\{R_g,\widehat \Phi_g\}_{g=1}^G\),
\begin{equation*}
\begin{aligned}
  \ell\bigg(V_0,\{R_g,\widehat \Phi_g\}_{g=1}^G\mid  \{Y_g,X_g,\widehat\Psi_g\}_{g=1}^G,&\{V_g\}_{g=1}^G\bigg)
=
C-\frac{1}{2}\sum_{g=1}^G\left[p\log\left|I_{k_g}+A_g^2\widehat \Phi_g\right| \right.\\
&\left.+\operatorname{tr}\left\{\widehat\Psi_g^{-1}(V_g-V_0R_g)A_g^2\left(I_{k_g}+A_g^2\widehat \Phi_g\right)^{-1}(V_g-V_0R_g)^\top\right\}\right].  
\end{aligned}
\end{equation*}

\noindent

\section{Derivation of the coordinate updates in Section \ref{subsec:eb}}

We derive the coordinate updates used to maximize the marginal likelihood in Section 2.4. From the expansion of the marginal likelihood, after replacing \(\Psi_g\) by the plug-in estimate \(\widehat\Psi_g\), the terms depending on \(V_0\), \(R_g\), and \(\Phi_g\) are
\begin{equation*}
\begin{aligned}
  \ell\bigg(V_0,\{R_g,\widehat \Phi_g\}_{g=1}^G\mid  \{Y_g,X_g,\widehat\Psi_g\}_{g=1}^G,&\{V_g\}_{g=1}^G\bigg)
=
C-\frac{1}{2}\sum_{g=1}^G\left[p\log\left|I_{k_g}+A_g^2\widehat \Phi_g\right| \right.\\
&\left.+\operatorname{tr}\left\{\widehat\Psi_g^{-1}(V_g-V_0R_g)B_g(V_g-V_0R_g)^\top\right\}\right].  
\end{aligned}
\end{equation*}
where
\begin{equation*}
B_g=A_g^2\left(I_{k_g}+A_g^2 \widehat \Phi_g\right)^{-1}, \quad (g=1, \dots, G)\end{equation*}
Since the \(A_g\)'s and \(\widehat \Phi_g\)'s are diagonal, the \(B_g\)'s are diagonal. For fixed \(\{\widehat \Phi_g\}_{g=1}^G\), the determinant term does not depend on \(V_0\) or \(\{R_g\}_{g=1}^G\). Hence maximizing the marginal likelihood with respect to \(V_0\) and \(\{R_g\}_{g=1}^G\) is equivalent to minimizing
\begin{equation*}
Q\left(V_0,\{R_g\}_{g=1}^G\right)
=
\sum_{g=1}^G\operatorname{tr}\left\{\widehat\Psi_g^{-1}(V_g-V_0R_g)B_g(V_g-V_0R_g)^\top\right\}.
\end{equation*}

\subsection{Update for \(V_0\)}
Fix \(\{R_g,\Phi_g\}_{g=1}^G\). Write \(\widehat\Psi_g=\operatorname{diag}(\widehat\sigma_{g1}^2,\ldots,\widehat\sigma_{gp}^2)\). Let \(v_{g, j}\in\mathbb R^{k_g}\) denote the \(j\)-th row of \(V_g\), and let \(v_{0,j}\in\mathbb R^{k_0}\) denote the \(j\)-th row of \(V_0\). Since \(\widehat\Psi_g\) is diagonal, the objective separates over rows:
\begin{equation*}
Q(V_0)=\sum_{j=1}^p Q_j(v_{0,j}), \qquad Q_j(v_{0,j})=\sum_{g=1}^G \frac{1}{\widehat\sigma_{gj}^2}\left(v_{g, j}-R_g^\top v_{0,j}\right)^\top B_g\left(v_{g, j}-R_g^\top v_{0,j}\right).
\end{equation*}
It is easy to see that the minimizer takes the form
\begin{equation*}
\widehat v_{0,j}
=
\left[\sum_{g=1}^G\frac{1}{\widehat\sigma_{gj}^2}R_gB_gR_g^\top\right]^{-1}
\left[\sum_{g=1}^G\frac{1}{\widehat\sigma_{gj}^2}R_gB_gv_{g, j}\right].
\end{equation*}
Substituting \(B_g=A_g^2(I_{k_g}+A_g^2\Phi_g)^{-1}\), we obtain
\begin{equation*}
\widehat v_{0,j}
=
\left[\sum_{g=1}^G\frac{1}{\widehat\sigma_{gj}^2}R_g\left(I_{k_g}+A_g^2\Phi_g\right)^{-1}A_g^2R_g^\top\right]^{-1}
\left[\sum_{g=1}^G\frac{1}{\widehat\sigma_{gj}^2}R_g\left(I_{k_g}+A_g^2\Phi_g\right)^{-1}A_g^2v_{g, j}\right],
\end{equation*}
where the order of \(A_g^2\) and \((I_{k_g}+A_g^2\Phi_g)^{-1}\) can be interchanged because both matrices are diagonal.

\subsection{Update for \(R_g\)}
We derive the unconstrained update for \(R_g\). Fix \(V_0\) and \(\widehat\Phi_g\), and define
\begin{equation*}
B_g=A_g^2\left(I_{k_g}+A_g^2\widehat\Phi_g\right)^{-1}.
\end{equation*}
Ignoring the semi-orthogonality constraint on \(R_g\), the group-specific objective is
\begin{equation*}
Q_g(R_g)=\operatorname{tr}\left\{\widehat\Psi_g^{-1}(V_g-V_0R_g)B_g(V_g-V_0R_g)^\top\right\}.
\end{equation*}
Expanding the trace gives
\begin{equation*}
Q_g(R_g)=\operatorname{tr}(\widehat\Psi_g^{-1}V_gB_gV_g^\top)-2\operatorname{tr}(R_g^\top V_0^\top \widehat\Psi_g^{-1}V_gB_g)+\operatorname{tr}(R_g^\top V_0^\top \widehat\Psi_g^{-1}V_0R_gB_g).
\end{equation*}
The first term does not depend on \(R_g\). Since \(\widehat\Psi_g^{-1}\) and \(B_g\) are symmetric, differentiating with respect to \(R_g\) gives
\begin{equation*}
\nabla_{R_g}Q_g(R_g)=-2V_0^\top \widehat\Psi_g^{-1}V_gB_g+2V_0^\top \widehat\Psi_g^{-1}V_0R_gB_g.
\end{equation*}
Setting the gradient equal to zero yields
\begin{equation*}
V_0^\top \widehat\Psi_g^{-1}V_0\,\widetilde R_g B_g=V_0^\top \widehat\Psi_g^{-1}V_gB_g.
\end{equation*}
The unconstrained minimizer is
\begin{equation*}
\widetilde R_g=\left(V_0^\top \widehat\Psi_g^{-1}V_0\right)^{-1}V_0^\top \widehat\Psi_g^{-1}V_g.
\end{equation*}
This is the unconstrained weighted least-squares minimizer. The semi-orthogonality constraint is then enforced separately by projecting \(\widetilde R_g\) onto \(\mathcal S_{k_0,k_g}\).

\section{Additional details about the numerical experiments}

The code to implement our methodology and replicate the experiments is available at the following \href{https://github.com/maurilorenzo/hierarchical_subspace_shrinkage_paper}{repository}

The \texttt{PC-partial pool} estimator shares a common low-dimensional subspace across groups while allowing the group-specific loading matrices and residual variances to differ. Let
$$
Y_{\mathrm{pool}} =
\begin{bmatrix}
Y_1^\top & \cdots & Y_G^\top
\end{bmatrix}^{\!\top}
$$
and let $\widehat V_{\mathrm{pool}}\in\mathbb{R}^{p\times r}$ contain the leading $r$ right singular vectors of $Y_{\mathrm{pool}}$, where $r$ is the dimension of the shared subspace. For each group $g$, we project the data onto this pooled subspace and compute its SVD:
$$
Y_g\widehat V_{\mathrm{pool}}
=
\widetilde U_g\widetilde A_g\widetilde W_g^\top + \widetilde U_{g \perp}\widetilde A_{g \perp}\widetilde W_{g \perp}^\top.
$$
where $\widetilde U_g\widetilde A_g\widetilde W_g^\top$ denote the $k_g$ leading components and 
$\widetilde U_{g \perp}\widetilde A_{g \perp}\widetilde W_{g \perp}^\top$ the remaining ones.  Using the $k_g$ leading projected components, we define the group-specific loading estimate as
$$
\widehat\Lambda_{g \mathrm{PP}}
=
\frac{1}{\sqrt{n_g}}\widehat V_{\mathrm{pool}}
\widetilde W_g\widetilde A_g.
$$
Thus, the span of $\widehat\Lambda_{g \mathrm{PP}}$ is contained in the common pooled subspace spanned by  $\widehat V_{\mathrm{pool}}$.
We estimate the latent factors via 
$$
\widehat F_{g \mathrm{PP}}
=
\sqrt{n_g} 
\widetilde U_g,
$$
and the residual error variances via the mean squared residual errors,

$$
\widehat\Psi_{g \mathrm{PP}}
=
\operatorname{Diag}\!\left\{
\frac{1}{n_g-r}
\operatorname{diag}\!\left[
(Y_g - \widehat F_{g \mathrm{PP}} \widehat\Lambda_{g \mathrm{PP}}^\top)^{\!\top}
(Y_g - \widehat F_{g \mathrm{PP}} \widehat\Lambda_{g\mathrm{PP}}^\top)
\right]\right\},
$$

and form

$$
\widehat\Sigma_{g \mathrm{PP}}
=
\widehat\Lambda_{g \mathrm{PP}}
(\widehat\Lambda_{g \mathrm{PP}})^\top+
\widehat\Psi_{g \mathrm{PP}}.
$$

Unlike \texttt{PC-pool}, which uses one loading matrix and one residual covariance estimated from the concatenated data, \texttt{PC-partial pool} estimates these quantities separately by group after imposing the shared-subspace constraint. We ran the MCMC algorithm for $5000$ iterations, discarding the first $2000$ as burn-in.

We implemented \texttt{SHARED-SUBSPACE} using the code available at \href{https://github.com/afranks86/shared-subspace}{https://github.com/afranks86/shared-subspace}.

\texttt{AJIVE} was implemented via the R package \texttt{CJIVE}.

\end{document}